\documentclass[aps, prl, superscriptaddress, twocolumn, showpacs, longbibliography, footnoography]{revtex4-1}
\usepackage{graphicx}
\usepackage{amsmath,amssymb,mathtools}
\usepackage{physics}
\usepackage{dsfont, bm} 
\usepackage[dvipsnames]{xcolor}
\usepackage[colorlinks, linkcolor=Red, citecolor=NavyBlue, urlcolor=NavyBlue]{hyperref}
\usepackage[all]{hypcap} 

\begin{document}

\title{Many-Body Non-Hermitian Skin Effect for Multipoles}
\author{Jacopo Gliozzi}
\affiliation{Department of Physics and Institute for Condensed Matter Theory, University of Illinois at Urbana-Champaign, Urbana, Illinois 61801-3080, USA}
\author{Giuseppe De Tomasi}
\affiliation{Department of Physics and Institute for Condensed Matter Theory, University of Illinois at Urbana-Champaign, Urbana, Illinois 61801-3080, USA}
\affiliation{Max Planck Institute for the Physics of Complex Systems, 01187 Dresden, Germany}
\author{Taylor L. Hughes}
\affiliation{Department of Physics and Institute for Condensed Matter Theory, University of Illinois at Urbana-Champaign, Urbana, Illinois 61801-3080, USA}


\begin{abstract}
In this work, we investigate the fate of the non-Hermitian skin effect in one-dimensional systems that conserve the dipole moment and higher moments of an associated global $\text{U}(1)$ charge. Motivated by field theoretical arguments and lattice model calculations, we demonstrate that the key feature of the non-Hermitian skin effect for $m$-pole conserving systems is the generation of an $(m+1)$th multipole moment. For example, in contrast to the conventional skin effect where charges are anomalously localized at one boundary, the dipole-conserving skin effect results in charges localized at both boundaries, in a configuration that generates an extremal \emph{quadrupole} moment. In addition, we explore the dynamical consequences of the $m$-pole skin effect, focusing on charge and entanglement propagation. Both numerically and analytically, we provide evidence that long-time steady-states have Fock-space localization and an area-law scaling of entanglement entropy, which serve as quantum indicators of the skin effect.
\end{abstract}

\maketitle

\textit{Introduction}.---
A central postulate of quantum mechanics is the unitarity of time evolution, which implies the conservation of probability during dynamics. However, interactions between a quantum system and its environment can induce dissipation, decoherence, or wavefunction collapse, destroying unitarity.
Recent advances in controlled experimental techniques that allow for single-qubit manipulation, along with the need to overcome decoherence in quantum devices, 
have led to increasing interest in non-unitary dynamics
as a paradigm to describe open and monitored quantum systems~\cite{Fisher_2023, Potter2022, Altman_2021, Chertkov2023, Ashida_2020}.
Relaxing unitarity allows for a rich phenomenology, ranging from novel topological phases~\cite{Leykam_2017, Kawabata2019topo, Kawabata2020, Borgnia2020, Ma2023, Okuma2023, Lavasani2021, Lavasani2021Topo} to phase transitions in the entanglement structure of non-equilibrium steady-states~\cite{Li2018, Skinner_19, Chan2019, Choi_2020, Bao1_2020, Gullans2020, Weinstein_2022}. 

One approach to non-unitary dynamics is through non-Hermitian Hamiltonians, which model quantum systems subject to dissipation~\cite{Feshbach, Rotter2009} or continuous measurements and postselection~\cite{Ueda1990, Dalibard1992, Dum1992, Plenio1998, Daley2014}. Remarkably, non-Hermitian couplings can lead to phenomena with no Hermitian counterpart, like the non-Hermitian skin effect, an anomalous localization of extensively many eigenstates at the boundary of a system~\cite{Lee2016,Yao2018, Review_non_Hermitian_Zhang_22, Lin2023}, which has been observed experimentally~\cite{Ananya_Ghatak_2020,Helbig2020,Zhang2021,Zhang12021, Shen2023}. While the boundary localization of particles is an essentially classical effect of non-reciprocal hoppings, it also impacts the spread of quantum correlations during time evolution. 
For example, the skin effect has been shown to hinder entanglement growth and can trigger a volume-to-area law transition in the entanglement entropy of long-time steady-states~\cite{Kawabata2023}. 
In addition, the skin effect challenges the traditional understanding of bulk-boundary correspondence in topological systems~\cite{Xiao2020,Zhang2022universal}. 

Previous studies of the skin effect have been mostly restricted to non-interacting Hamiltonians with $\text{U}(1)$ particle-number symmetry. Moreover, the few extensions to interacting systems have relied on the existence of a non-interacting limit~\cite{Zhang2020interact, Liu2020interact, Kawabata2022, Zhang2022}. In this work, we study the non-Hermitian skin effect in intrinsically interacting systems that conserve multipole moments of a $\text{U}(1)$ charge.
Such conservation laws constrain particle motion and can dramatically affect dynamics~\cite{Pai_2019, Khemani_2020, Feldmaier_2020, Rakovszky2020, Sala2020, Feldmeier_2021, Burchards_2022, Feldmeier_2022, Gliozzi_2023, Zechmann_2023,Ogunnaike_2023}.
For instance, in systems that conserve the total dipole moment, isolated charges cannot move, leading to anomalously slow charge transport and ergodicity breaking.

\begin{figure}[t!]
\includegraphics[width=\linewidth]{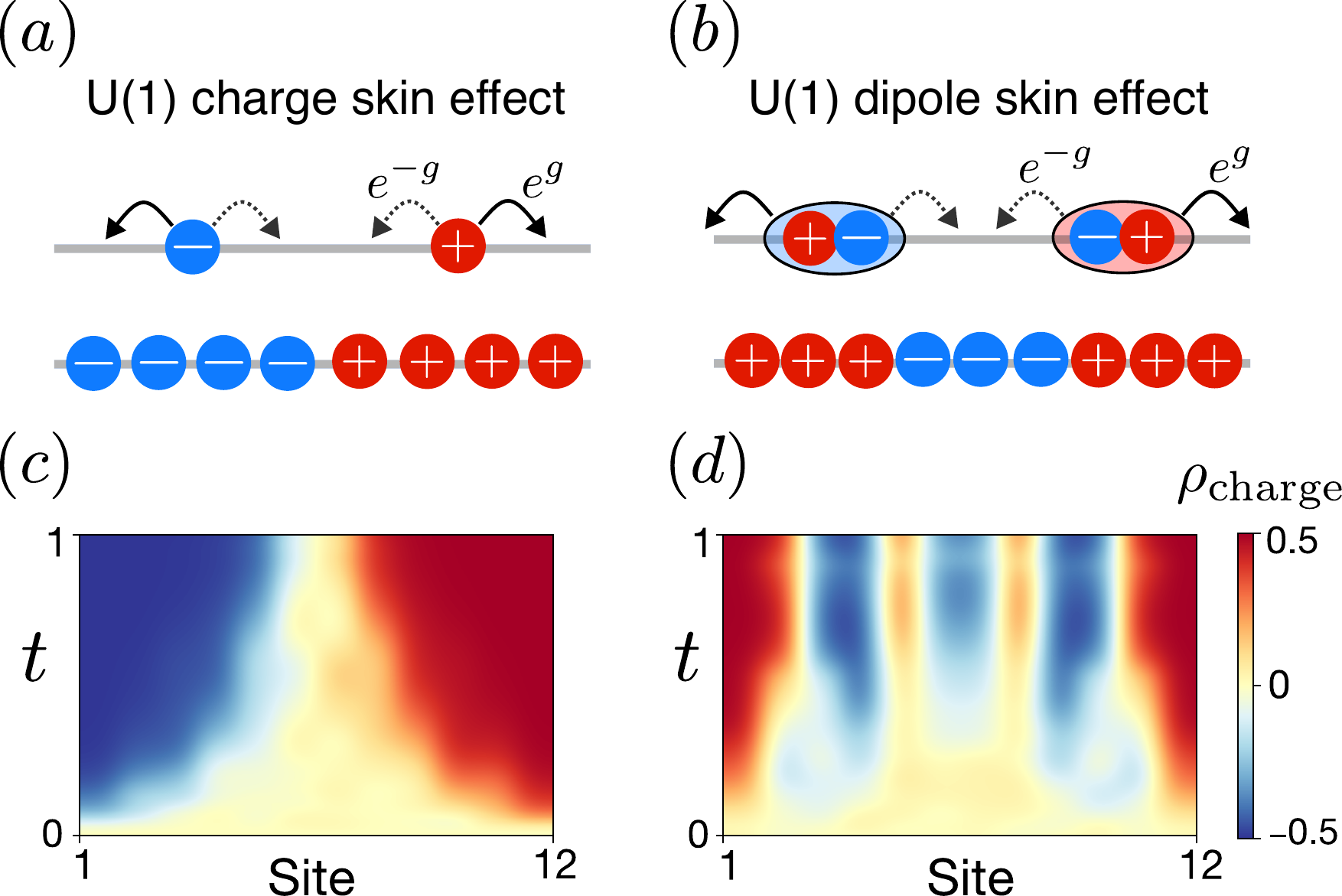}
\caption{
(a-b) Schematic of many-body non-Hermitian charge and dipole skin effect, which generate a dipole and quadrupole moment, respectively.
(c-d) Time evolution of the local charge density $\rho_\text{charge}$ from a random uniform state for (c) charge-conserving and (d) dipole-conserving Hamiltonians with a non-Hermitian imbalanced kinetic term.
}\label{fig:1}
\end{figure}

The interplay between multipole conservation laws, which restrict particle motion, and non-reciprocal hoppings, which pump particles to a boundary through the non-Hermitian skin effect, creates a complex scenario for charge transport. For example, charge flows to a single boundary with only $\text{U}(1)$ symmetry, but in dipole-conserving systems it must flow symmetrically outwards to both boundaries (see Fig.~\ref{fig:1}).
Motivated by the hierarchical structure of multipoles, we define the $m$-pole skin effect in terms of an $(m+1)$-pole moment in the charge distribution of eigenstates. We also investigate the dynamical consequences of this generalized skin effect by examining charge and entanglement propagation after a quench. In particular, we provide analytical and numerical evidence that the $m$-pole skin effect leads to area-law entanglement entropy in long-time steady-states. Importantly, our definition encompasses the standard ``charge'' skin effect, and helps identify
experimentally accessible probes in systems with strong interactions. 

\textit{Charge skin effect}.---
Before discussing higher multipoles, let us first revisit the non-Hermitian skin effect in the context of $\text{U}(1)$ charge conservation. 
We focus on one-dimensional fermionic chains of length $L$. The paradigmatic non-Hermitian model for the skin effect is the Hatano-Nelson Hamiltonian~\cite{Hatano1996, Hatano1997}:
\begin{equation}\label{eq:hatano}
\begin{gathered}
    H_\text{HN}(g) = - \sum_j (e^{g} c^\dag_j c_{j+1} + e^{-g} c^\dag_{j+1} c_j),
\end{gathered}
\end{equation}
where $c^\dag_j$ is the fermionic creation operator at site $j,$ and $g$ parameterizes the left-right imbalance in charge hopping. 
With periodic boundary conditions (PBCs), the single-particle spectrum of Eq.~\eqref{eq:hatano} winds around $E=0$ in the complex plane, and the eigenstates are extended plane waves. In contrast, for open boundary conditions (OBCs) the spectrum is real-valued, and all single-particle eigenstates are exponentially localized at one boundary. This stark difference between PBC and OBC spectra is the hallmark of the non-Hermitian skin effect for non-interacting systems~\cite{Lee2016, Yao2018}. 

To extend this definition to interacting systems, we first need a many-body diagnostic of the skin effect. Unlike the single-particle case, 
an extensive number of particles cannot be exponentially boundary-localized in a many-body state because of the Pauli exclusion principle, which forbids fermions from clustering at the same site~\cite{Lee2020,Liu2020}. 
Nevertheless, features of the single-particle phenomenology persist. For instance, the $N$-particle spectrum of Eq.~\eqref{eq:hatano} winds in the complex plane when the PBCs are twisted by a phase. In OBCs, the spectrum is still real-valued, and the eigenstates exhibit a left-right imbalance of charge.
This asymmetry in the charge density of open eigenstates has been interpreted as a many-body extension of the skin effect~\cite{Alsallom2022}.

While such diagnostics have been used for the Hatano-Nelson model with Hubbard interactions~\cite{Kawabata2022, Zhang2022}, we aim for a definition that applies even in the absence of a non-interacting limit.
Independently of particle number or interactions, the skin effect arises from the imbalanced kinetic term, which preferentially moves electrons to one edge and holes to the other. The resulting charge distribution, depicted in Fig.~\ref{fig:1}(a), exhibits an electric \emph{dipole} moment, which is defined by 
\begin{equation}\label{eq:dipole}
\begin{gathered}
    P^{(1)} = \sum_{j=1}^L j \rho_j ,
\end{gathered}
\end{equation}
where $\rho_j$ is the deviation of the local density at site $j$, $n_j = c^\dag_j c_j$, from the average density per site, $\bar{n} = N/L$. We then define the non-Hermitian skin effect as the presence of an extensive dipole moment in the $N$-particle OBC eigenstates, formalizing the previous notion of a left-right asymmetry in density. Importantly, the dipole moment arises both in the distribution of eigenstates and dynamically from time evolution, as shown in Fig.~\ref{fig:1}(c). We will see that this definition directly generalizes to systems that conserve arbitrary $\text{U}(1)$ multipole moments.

For the Hatano-Nelson model, the relation between imbalanced charge-hopping and dipole moment can be understood by applying an ``imaginary gauge transformation''~\cite{Hatano1996, Hatano1997}:
\begin{equation}\label{eq:similarity}
\begin{gathered}
c_j^\dag \mapsto R^{-1} c_j^\dag R = e^{g j} c_j^\dag, \quad R = e^{-g P^{(1)}},
\end{gathered}
\end{equation}
where $P^{(1)}$ is the dipole moment in Eq.~\eqref{eq:dipole}~\footnote{The corresponding transformation of annihilation operators is $c_j \rightarrow R^{-1} c_j R = e^{-gj} c_j$.}.
In a chain with OBCs, this is a similarity transformation that maps $H_\text{HN}(g)$ to $H_\text{HN}(0)$, a \emph{Hermitian} Hamiltonian without imbalance. An immediate consequence of the map is that the OBC spectrum of $H_\text{HN}(g)$ must be real. Furthermore, single-particle eigenstates of the Hermitian $H_\text{HN}(0)$ are standing wave-like, so applying the map in reverse yields eigenstates of the original Hamiltonian that are exponentially localized at an edge~\footnote{A single-particle state $\ket{\psi_{0\alpha}}=\sum_j \alpha_j c^\dag_j \ket{0}$ is mapped back to $\ket{\psi_{g\alpha}}=\sum_j e^{gj} \, \alpha_j  c^\dag_j \ket{0}$, which is boundary-localized as long as $\alpha_j$ is not fine-tuned to counteract the exponential dependence}.
Regardless of particle number, the operator $R$ tends to project onto states that maximize or minimize the dipole operator (depending on the sign of $g$).
Hence, while a generic many-body eigenstate $\ket{\psi_0}$ of $H_\text{HN}(0)$ does not exhibit charge imbalance, the corresponding eigenstate of $H_\text{HN}(g)$, $\ket{\psi_g} = R^{-1} \ket{\psi_0}$, necessarily has an extremal dipole moment~\footnote{The many-body eigenstates of $H_\text{HN}(g)$ are Slater determinants of the single-particle states $\ket{E(g)}= R^{-1} \ket{E(0)}$.}.

Although the map in Eq.~\eqref{eq:similarity} is applicable only to Hamiltonians with a real spectrum, imbalanced charge hopping also induces a dipole moment in more general models.
We can argue this heuristically using the effective field theory approach for non-Hermitian systems developed in Ref.~\onlinecite{Kawabata2021}. There, the authors couple the kinetic term of a non-Hermitian Hamiltonian to a background $\text{U}(1)$ gauge field through a Peierls factor $c^\dag_j e^{i A_{j, j+1}} c_{j+1}$, and derive a response action for the vector potential $A_x$ in the continuum limit,
\begin{equation}
\begin{gathered}
    \label{eq:response_action}
    S_E = W(E) \int dx \, A_x,
\end{gathered}
\end{equation}
where $E$ is a reference energy, and $W(E)$ is the winding of the PBC spectrum around $E$.
This effective action encodes the response of quasiparticle modes of energy $E$, and though $W(E)$ depends on the underlying Hamiltonian, the form of the action is fixed by $\text{U}(1)$ symmetry.
The main observable associated with Eq.~\eqref{eq:response_action} is a persistent unidirectional current: $j^x(E) \coloneqq \delta S_E[A_x]/\delta A_x = W(E)$~\footnote{With open boundaries, the effective action must include boundary degrees of freedom to preserve gauge invariance. Physically, these are external sources and sinks of charge that allow a steady-state current to exist.}. Expressing the current in terms of the total dipole moment, $j^x = \partial_t P^{(1)}$, it follows that the physical response of a non-Hermitian charge-conserving system is the generation of a dipole moment when $W(E) \neq 0$.

\begin{figure}[t]
\hspace{-0.7cm} \includegraphics[width=\linewidth]{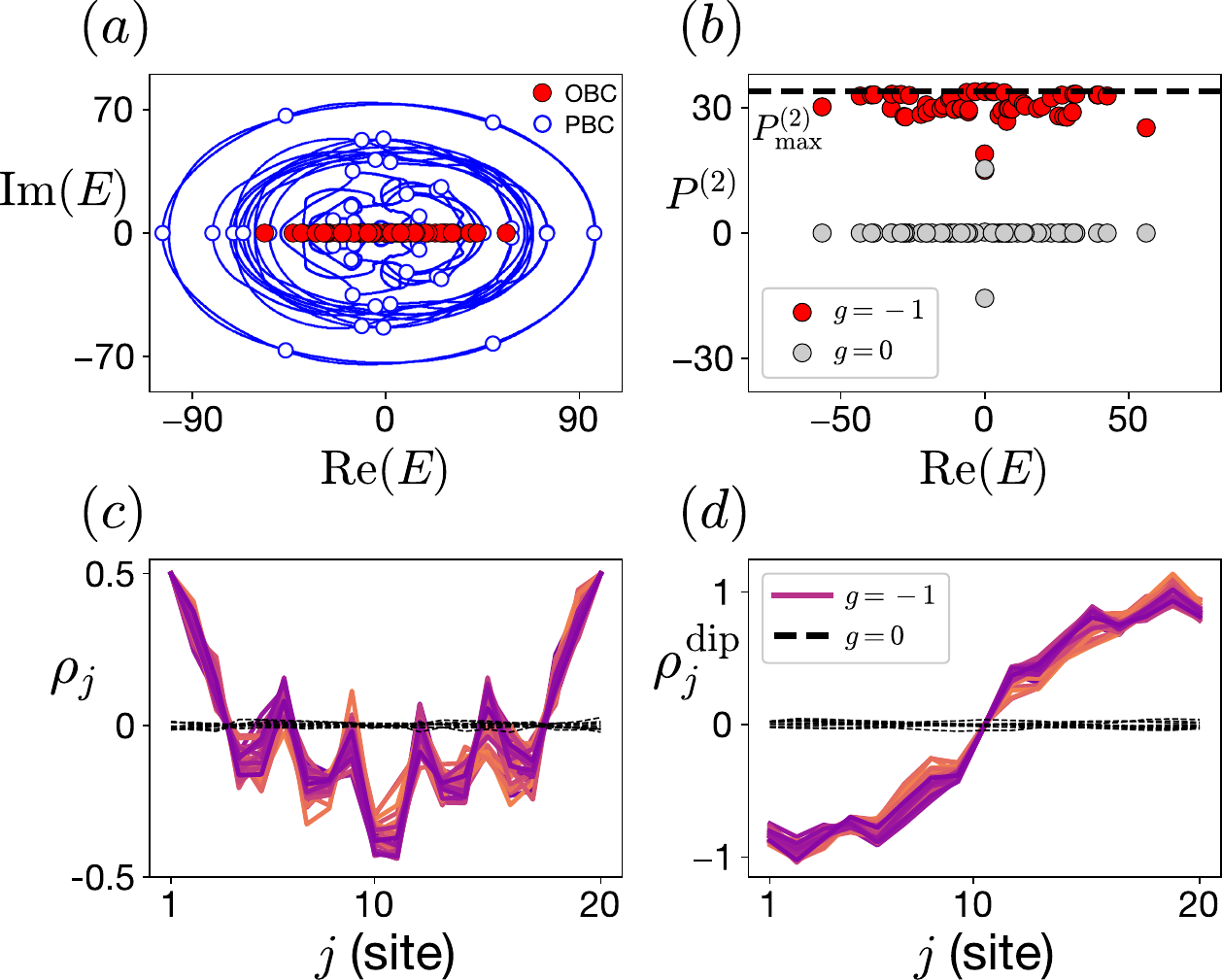}
\caption{
(a) Many-body energy spectrum of $H_\text{dip}$ at half-filling and $P^{(1)} = 0$ ($L=12, \, g \coloneqq g_1=2g_2/3=-1, \, t=0.6$). For OBCs (red dots), the spectrum is real, while for PBCs (empty dots) the spectrum is complex and winds as the boundary conditions are twisted (blue lines). (b) Quadrupole moment $P^{(2)}$ of right eigenstates of the open chain with (red) and without (grey) imbalance. All eigenstates with $g\neq0$ exhibit near-maximal quadrupole moment (dashed line), a signature of the dipole skin effect. (c) Local charge density and (d) local dipole density of 30 mid-spectrum states for $L=20$ and OBCs. The charge distribution exhibits a quadrupole moment, in contrast to that of the analogous Hermitian model (black dashed lines).
}
\label{fig:2}
\end{figure}

\textit{Multipole skin effect}.---
Having defined the charge skin effect in terms of the dipole operator, we now turn to the higher multipole case. In particular, we will show that the dipole-conserving skin effect is associated with a \emph{quadrupole moment} in the charge distribution. This generalizes to a hierarchical structure in which the $m$-pole skin effect is defined through the $(m+1)$-pole moment. 
To demonstrate, we begin by considering an analytically tractable model that conserves both $\text{U}(1)$ charge $N$ and dipole moment $P^{(1)}$~\footnote{The dipole moment is only conserved modulo the system size $L$ in periodic boundary conditions.},
\begin{equation}\label{eq:h4}
\begin{gathered}
H_4(g) = -\sum_j \left(e^{g}\, c^\dag_j c_{j+1} c_{j+2} c^\dag_{j+3}  + e^{-g} \, c_j c^\dag_{j+1} c^\dag_{j+2} c_{j+3}  \right).
\end{gathered}
\end{equation}
This Hamiltonian generalizes the Hatano-Nelson model to imbalanced dipole-hopping terms, where $d^\dag_j \coloneqq c^\dag_j c_{j+1}$ creates a local dipole at site $j$ (see Fig.~\ref{fig:1}(b)). 
Like the Hatano-Nelson model, the many-body spectrum of $H_4(g)$ is complex for PBCs and real for OBCs.

Taking $L$ to be divisible by four for simplicity, we consider the model at half-filling $N=L/2$ and fixed dipole moment $P^{(1)} = 0$, i.e., the largest symmetry sector. Because $H_4$ exhibits strong Hilbert space fragmentation, this symmetry sector itself splits into exponentially many disconnected Krylov subsectors~\cite{Sala2020,Moudgalya2021, Shen2022}. The largest of these subsectors is integrable and can be mapped to a free-fermion chain of length $L/2$~\cite{Moudgalya2021}.
The mapping proceeds by grouping sites $2j-1$ and $2j$ of the original chain into a new composite site, which hosts four degrees of freedom associated with the original site occupations: $\ket{00}, \ket{01}, \ket{10}, \ket{11}$. Composite sites are exactly singly-occupied in the largest subsector, and the allowed states $\ket{10}$ and $\ket{01}$ can be interpreted as a local dipole and anti-dipole, respectively. The dipole creation and annihilation operators on odd sites, $d^\dag_{2j-1}$ and $d_{2j-1}$, thus act as genuine fermionic operators on this subspace.

Only the odd terms in Eq.~\eqref{eq:h4} act nontrivially in the subsector, so the resulting effective Hamiltonian can be written as $H_4^\text{sub} = \sum_k (e^{g} d^\dag_k d_{k+1} + e^{-g} d_k d^\dag_{k+1})
$, where $k$ labels the composite sites.
This is just the Hatano-Nelson model, though with the fermions interpreted as local dipoles instead of charges. As a result, the non-Hermitian skin effect is manifest as a dipole moment of \emph{dipoles}, and therefore a quadrupole moment of charges. The buildup of dipoles (anti-dipoles) at the right (left) end of an open chain is depicted in Fig.~\ref{fig:1}(b), and generates an extensive quadrupole moment:
\begin{equation}\label{eq:quadrupole}
\begin{gathered}
P^{(2)} = \sum_{j=1}^L j^2 \rho_j.
\end{gathered}
\end{equation}
In contrast to the ordinary skin effect, there is no asymmetry in the charge distribution of OBC eigenstates, and particles are localized at both boundaries. 

A few comments about the generality of this model are in order. Although the quadrupolar response is particularly clear in the example above, the mapping between $H_4(g)$ and a free-fermion chain holds only in a few integrable subsectors. Furthermore, the strong Hilbert space fragmentation precludes ergodicity, making it difficult to observe the skin effect dynamically. Finally, because the dipole operator $P^{(1)}$ does not commute with translations~\cite{Gromov2020}, local dipoles themselves are not always well-defined. For example, a sequence of alternating positive and negative charges can be viewed as a chain of either local dipoles \emph{or} anti-dipoles.

To circumvent these restrictions, we consider a more general dipole-conserving Hamiltonian: 
\begin{equation}\label{eq:h4h5}
\begin{aligned}
H_\text{dip}(g_1, g_2) &= -\sum_{j} \left(h^{(2)}_j(g_1) + t \, h^{(3)}_j(g_2)\right), \\
h^{(k)}(g) &= e^{g} c^\dag_j c_{j+1} c_{j+k} c^\dag_{j+k+1} + e^{-g} \times \text{h.c.},
\end{aligned}
\end{equation}
which reduces to Eq.~\eqref{eq:h4} for $t=0$, but otherwise contains longer range dipole-hopping terms that prevent strong fragmentation~\cite{Sala2020}. Once again, we fix $N=L/2$ and $P^{(1)}=0$. The resulting PBC spectrum winds around the complex plane when the boundary conditions are twisted, as shown via exact diagonalization in Fig.~\ref{fig:2}(a). 
Many-body windings generically diverge in the thermodynamic limit~\footnote{{For example, the single-particle Hatano-Nelson spectrum in PBCs becomes a circle in the complex plane as $L\rightarrow \infty$. Many-body energies are sums of single-particle energies, so the many-body PBC spectrum is a dense disk.}}, {obscuring the topology of the skin effect,} so we focus on OBCs~\cite{Kawabata2022}.
The model in Eq.~\eqref{eq:h4h5} is not integrable and cannot be mapped to a free-fermion Hamiltonian.
Nevertheless, for the particular choice of parameters $g_2 = 3g_1/2$, there exists a similarity transformation analogous to Eq.~\eqref{eq:similarity} that maps $H_\text{dip}(g, 3g/2)$ to the Hermitian Hamiltonian $H_\text{dip}(0,0)$:
\begin{equation}\label{eq:similarity_dip}
\begin{gathered}
c_j^\dag \mapsto R^{-1} c_j^\dag R = e^{g j^2/4} c_j^\dag, \quad R = e^{-g P^{(2)}/4}.
\end{gathered}
\end{equation}
The spectrum of $H_\text{dip}(g, 3g/2)$ is therefore real, and the eigenstates of the non-Hermitian model are related to the Hermitian eigenstates by $R^{-1}$, which manifestly projects onto states with extremal quadrupole moment $P^{(2)}$. We claim that this is a fundamental signature of the ``dipole skin effect'', which, as we have seen, arises in generic dipole-conserving Hamiltonians with non-reciprocal hoppings. Motivated by the charge- and dipole-conserving cases, we can generalize the definition to systems that conserve up to $m$th multipole moments: the non-Hermitian $m$-pole skin effect is marked by the generation of an extensive $(m+1)$-pole moment in the open-boundary eigenstates~\footnote{Multipole moments are origin-dependent when lower moments do not vanish, but for any choice of origin, the $m$-pole skin effect always tends to extremize the $(m+1)$-pole moment.}. 

We confirm the dipole skin effect through exact diagonalization of $H_\text{dip}(g_1,g_2)$: the real OBC spectrum is shown in Fig.~\ref{fig:2}(a) for $L=12$, while the near-maximal quadrupole moments of the eigenstates are plotted in Fig.~\ref{fig:2}(b). The charge density profiles of a few mid-spectrum eigenstates for $L=20$ are shown in Fig.~\ref{fig:2}(c): here the quadrupole moment is manifest as a buildup of positive charge at the system boundaries and negative charge in the middle. Following Refs.~\cite{Moudgalya2021statistics, Zechmann2023}, we can also define a bounded local dipole density $\rho^\text{dip}_j = - \sum_{k \leq j} \rho_k$~\footnote{This comes from expressing the dipole operator as $P^{(1)} = -\sum_{j=1}^L (L-j) \rho_j$ (since $\sum_j \rho_j=0$ by definition) and then rewriting the sum as $ -\sum_{j=1}^{L-1} \sum_{k=1}^j \rho_j$}, which we plot for the same eigenstates in Fig.~\ref{fig:2}(d). The imbalance in dipole hopping yields a dipole moment of dipoles, i.e., a quadrupole moment of the charge, which should be contrasted with the Hermitian case, where $\rho_j$ and $\rho^{\text{dip}}_j$ are uniformly distributed across the sample (dashed lines). 

The similarity transformation in Eq.~\eqref{eq:similarity_dip} helps prove the existence of the skin effect, but generic non-reciprocal Hamiltonians with complex OBC spectra exhibit the same phenomenology. Indeed, we have checked numerically that our results are qualitatively unchanged for arbitrary values of $g_1$ and $g_2$ \footnote{{See Supplemental Material for additional numerics on systems with arbitrary $g_1$ and $g_2$ and systems that violate dipole conservation}}. This stability can be justified by an effective field theory that we propose for dipolar response. The background gauge field for dipole symmetry, $A_{xx}$, is a (one-component) rank-2 tensor and gives the Peierls phase associated with a dipole-hopping process~\cite{Pretko2017}. By analogy with Eq.~\eqref{eq:response_action}, we expect a response action of $S_E =  \int dx j^{xx} A_{xx}$ for generic dipole-conserving non-reciprocal systems, whose primary observable is a dipole current $j^{xx} = \partial_t P^{(2)}$ that generates a quadrupole moment in open boundaries~\cite{Hughes2021}. 
\begin{figure}[t!]
\label{fig:3}
\includegraphics[width=\linewidth]{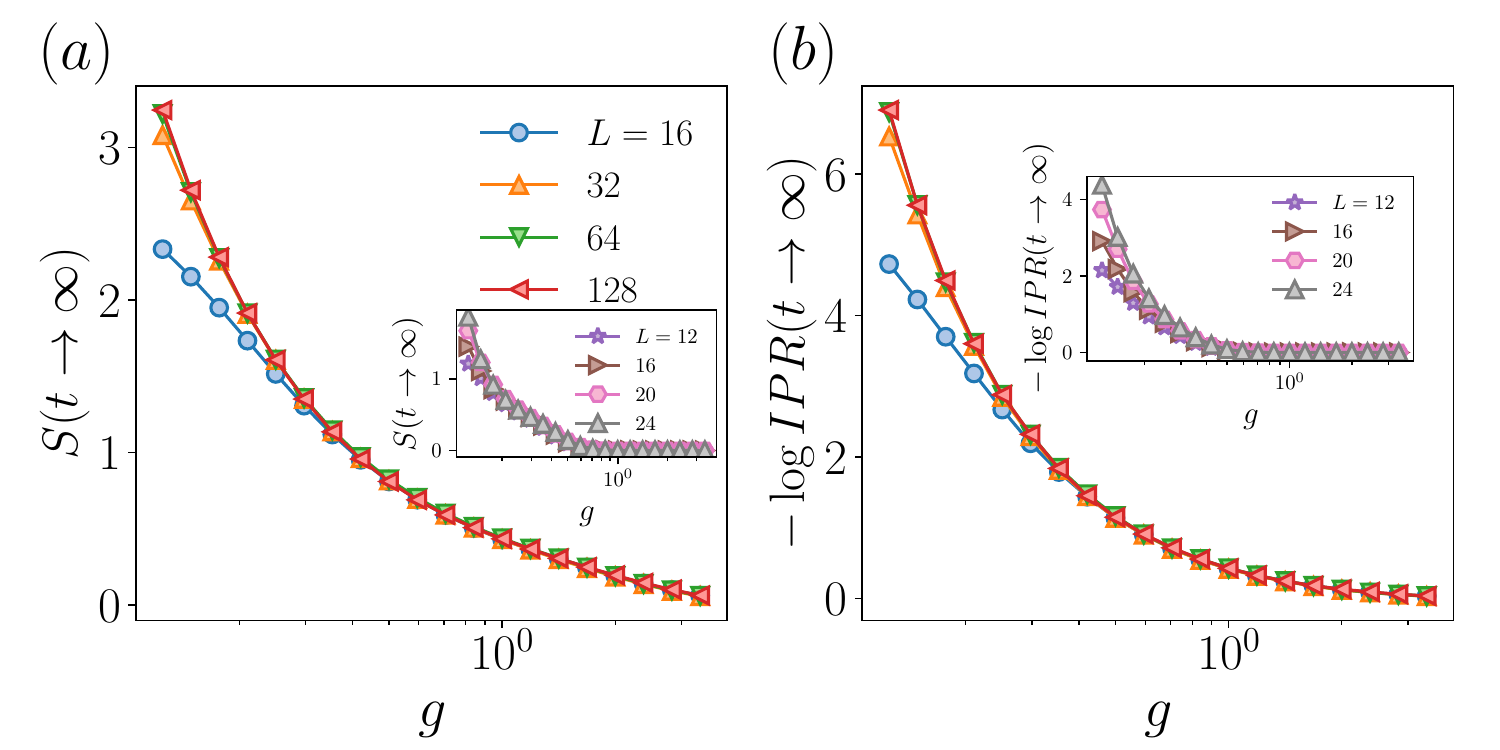}
\caption{
Dynamical probes of the late-time steady-state after evolution with charge-conserving (inset: dipole-conserving) non-Hermitian Hamiltonian with non-reciprocal hopping imbalance $g$, {specifically $H_\text{HN}(g)$ ($H_\text{dip}(g, 3g/2)$).} (a) The half-chain von Neumann entanglement entropy exhibits area-law scaling. (b) The inverse participation ratio converges with system size, indicating Fock-space localization. Both quantities are averaged over random initial product states.
}
\end{figure}

\textit{Dynamical signatures}.---Having defined and examined the skin effect in systems that conserve multipole moments, we now focus on its implications for dynamics. 
Dynamical probes are especially important because they can be implemented in experimental platforms to characterize the skin effect. Concretely, we study the state of a system at time $t$ undergoing dynamics modeled by a non-Hermitian Hamiltonian $H$:
\begin{equation}
\begin{gathered}
\label{eq:evolution}
    |\psi(t)\rangle =  \frac{e^{-i t H}|\psi(0)\rangle}{\| e^{-i t H}|\psi(0)\rangle \|}.
\end{gathered}
\end{equation}

We have already discussed the dynamics of charge: the $m$-pole skin effect causes charged particles to flow and generate an $(m+1)$-pole moment. Now, we turn to the evolution of quantum correlations, focusing on the spread of entanglement. Entanglement is quantified by the von Neumann entropy $S(t) = -\text{Tr}[\rho_{L/2} \log{\rho_{L/2}}]$, where $\rho_{L/2}$ is the reduced density matrix of the evolved state $|\psi(t)\rangle$ for a spatial cut that divides the system in half. Generic closed quantum systems, such as $H_{\text{dip}}(0,0)$, thermalize under their own time evolution. It follows that generic states evolve to have volume-law entanglement, $S\sim {O}(L)$. This behavior should be distinguished from area-law entanglement, $S\sim O(1)$, which is typically found only for ground states.

Besides entanglement, we also investigate the spread of $|\psi(t)\rangle$ in Fock space through the inverse participation ratio, $IPR_2(t) = \sum_{\overline{n}} | \langle \overline{n}| \psi(t)\rangle |^4$, where $\{|\overline{n}\rangle\}$ is the many-body computational basis. The evolved state spans the entire Hilbert space if $IPR_2\sim 2^{-D_2L}$ with fractal dimension $D_2 = 1$, and is localized if $D_2=0$. Let us consider a generic $m$-pole conserving Hamiltonian with asymmetric hopping such that a similarity transformation $R^{-1} H^{m}(g) R = H^{m}(0)$ exists, with $R = e^{-\beta(g) P^{(m+1)}}=e^{- \beta(g) \sum_j j^{m+1} \rho_j}$. Starting from a random product state in the computational basis, $\ket{\psi(0)} = \prod_{\{s_j\}} c_{s_j}^\dagger \ket{0}$, we can use the transformation to write 
\begin{equation}
\begin{gathered}
IPR_2(t) = \frac{\sum_{\underline{n}} e^{-4\beta \sum_j j^{m+1} n_j} |\langle \underline{n} | e^{-i t H^{(m)}(0)} |\psi(0)\rangle |^4}{ (\sum_{\underline{n}} e^{-2\beta \sum_j j^{m+1} n_j} |\langle \underline{n} | e^{-i t H^{(m)}(0)} |\psi(0)\rangle |^2 )^2}.
\end{gathered}
\end{equation}
Assuming that the corresponding Hermitian evolution reaches a generic state, e.g., via thermalization, such that $|\langle \underline{n} | e^{-i t H^{(m)}(0)} |\psi(0)\rangle |^2 \sim 2^{-L}$ for large $t$, we obtain 
\begin{equation}
\begin{gathered}
|{IPR_2(t\rightarrow \infty)}| \approx \prod_{j=1}^L \frac{(1+e^{-4\beta  j^{m+1}}) }{(1+e^{-2\beta j^{m+1}})^2}.
\end{gathered}
\end{equation}
This expression is bounded by
$|\log{IPR_2}| \le \sum_{j=1}^L \left |\log{\left (1 - \frac{2 e^{-2 \beta j^{m+1}}}{(1+e^{-2\beta j^{m+1}})^2} \right )}\right |
$, which converges as $L\rightarrow \infty$, implying localization in Fock space ($D_2\rightarrow 0$). This should be viewed as a generalization of the edge localization that results from the single-particle skin effect~\footnote{{The strength of the localization can also be seen using the similarity transformation, which breaks up the evolution into a unitary piece, which generically yields a completely delocalized state, and the similarity map, $e^{-\beta(g) P^{(m+1)}}$, which then localizes this state in the corner of the Fock space with maximal $(m+1)$-pole moment. Consequently, $\beta(g) \sim g$ can be interpreted as a localization length.}} . Although fermion repulsion prevents many-body states from localizing at the physical boundary, the $m$-pole skin effect extremizes the $(m+1)$-pole moment of any initial state, driving time evolution to a highly restricted region of the Fock space. Finally, Fock-space localization implies that the entanglement entropy of the steady-state has area-law scaling~\cite{GDT_2020}.

We test our conjecture by performing exact numerical simulations of charge- and dipole-conserving time evolution {using free fermion techniques for the non-interacting case and
Chebyshev polynomial methods for the interacting case}~\cite{Weisse2006, Soumya_17}. 
Fig.~\ref{fig:3}(a-b) shows $S(t\rightarrow \infty)$ and $IPR_2(t\rightarrow \infty)$ for the $\text{U}(1)$ charge skin effect using $H_{\text{HN}}(g)$ in Eq.~\eqref{eq:hatano} (as a function of non-reciprocity $g$ and for various system sizes). In the insets of Fig.~\ref{fig:3}(a-b), we consider the dipole skin effect with $H_4(g)$ in Eq.~\eqref{eq:h4}. In agreement with our analysis, we find that the final states exhibit area-law scaling for the entanglement entropy and are localized in Fock space.

\textit{Conclusions}---
Non-reciprocal hopping leads to the localization of charge at one boundary of a 1D system. On the other hand, $m$-pole conservation imposes strong constraints on charge transport. The competition between these two effects gives rise to a new class of non-Hermitian skin effects in intrinsically interacting systems.
In this work, we studied these generalized skin effects using lattice models, field theory arguments, and numerics. Our central result, that eigenstates exhibiting the $m$-pole skin effect have an $(m+1)$-pole moment, naturally implies a definition of the ordinary skin effect beyond the non-interacting regime.

Furthermore, this definition provides an experimentally accessible signature of the $m$-pole skin effect, which may then be observed in ultra-cold atom platforms. For example, non-reciprocal charge hopping can be engineered by coupling an optical lattice to external dissipation~\cite{Gong2018}. Placing this lattice in a strongly tilted external field then effectively imposes dipole-conservation~\cite{Guardado-Sanchez2020,Scherg2021, Kohlert2023}, and the ensuing dynamics is governed by the dipole skin effect. Looking beyond local densities of conserved charges, we have also provided a truly quantum characterization of the skin effect, showing that it results in area-law entanglement entropy in steady-states.

We have seen an intimate connection between the symmetries of a system and its non-Hermitian dynamics. To this end, it would be interesting to consider models with other types of symmetries, like {spinful models with time-reversal symmetry, or higher-form and subsystem symmetries}, as well as patterns of symmetry-breaking. It may also be possible to extend the $m$-pole skin effect to higher dimensions and connect it to many-body topological invariants, both fruitful directions for future research. 
\\

\begin{acknowledgments}
G.D.T. acknowledges the support from the EPiQS Program of the Gordon and Betty Moore Foundation. T.L.H. and J.G. acknowledge support from the U.S. Office of Naval Research (ONR) Multidisciplinary University Research Initiative (MURI) under Grant No. N00014-20-1-2325 on Robust Photonic Materials with Higher-Order Topological Protection. {G.D.T. acknowledges the hospitality of MPIPKS Dresden where part of the work was done.}
\end{acknowledgments}

\bibliographystyle{apsrev4-1}
\bibliography{main.bib}

\setcounter{figure}{0}
\renewcommand{\thefigure}{S\arabic{figure}}
\setcounter{equation}{0}
\renewcommand{\theequation}{S\arabic{equation}}

\clearpage
\begin{center}
    \large\textbf{Supplementary Material}
\end{center}

\noindent
{\bf SM 1: Dipole skin effect in the absence of a similarity transformation to a Hermitian system}
\\
\indent
In the main text, much of our investigation of the dipole skin effect (and its higher multipole relatives) has been limited to non-Hermitian systems with a real spectrum in open boundary conditions (OBCs). In these cases, there exists a similarity transformation that maps the non-Hermitian Hamiltonian to a Hermitian one, greatly facilitating the analysis. Nevertheless, we heuristically argued using a response field theory that the dipole skin effect is also present in more general non-Hermitian Hamiltonians with a complex spectrum. Here, we provide some further numerical evidence for this claim.

\begin{figure}[b!]
\label{fig:S1}\hspace{-1cm}
\includegraphics[width=0.9\linewidth]{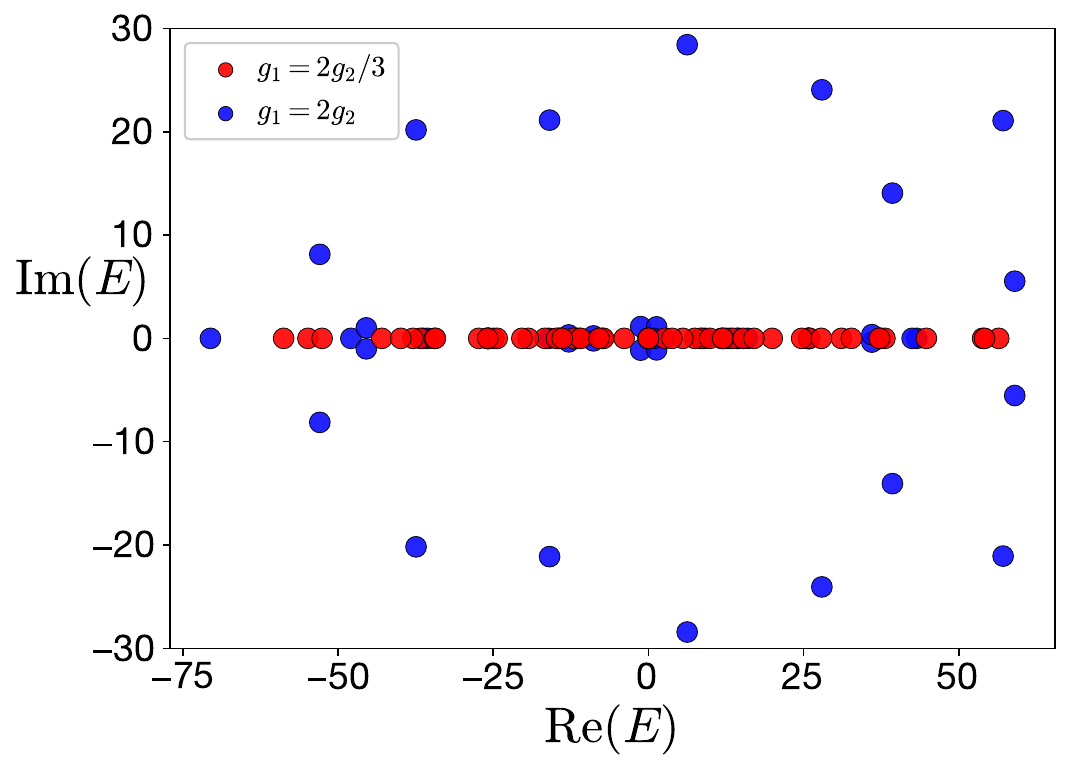}
\caption{Energies in OBCs ($L=12$) for $H_\text{dip}(g_1, g_2)$ for cases where a similarity transformation exists (red dots, $g_1 = 2/3, g_2 = 1$) and does not exist (blue dots, $g_1 = 2, g_2 = 1$). The former are always real, while the latter are complex.} 

\end{figure}We focus on the dipole-conserving non-reciprocal Hamiltonian that we introduced in Eq. (7) of the main text, which we reproduce here for convenience:
\begin{equation}\label{eq:h4h5}
\begin{aligned}
H_\text{dip}(g_1, g_2) &= -\sum_{j} \left(h^{(2)}_j(g_1) + t \, h^{(3)}_j(g_2)\right), \\
h^{(k)}(g) &= e^{g} c^\dag_j c_{j+1} c_{j+k} c^\dag_{j+k+1} + e^{-g} \times \text{h.c.},
\end{aligned}
\end{equation}
We henceforth fix $t=1$, and work in the symmetry sector with zero dipole moment and magnetization. For $g_1 = 2g_2/3$, this Hamiltonian has a real energies in open boundary conditions. Consequently, the similarity transformation performed by $R = e^{-g P^{(2)}/4}$, where $P^{(2)}$ is the total quadrupole moment, maps this Hamiltonian to a Hermitian model. 

\begin{figure}[h!]
\label{fig:S2}
\includegraphics[width=0.85\linewidth]{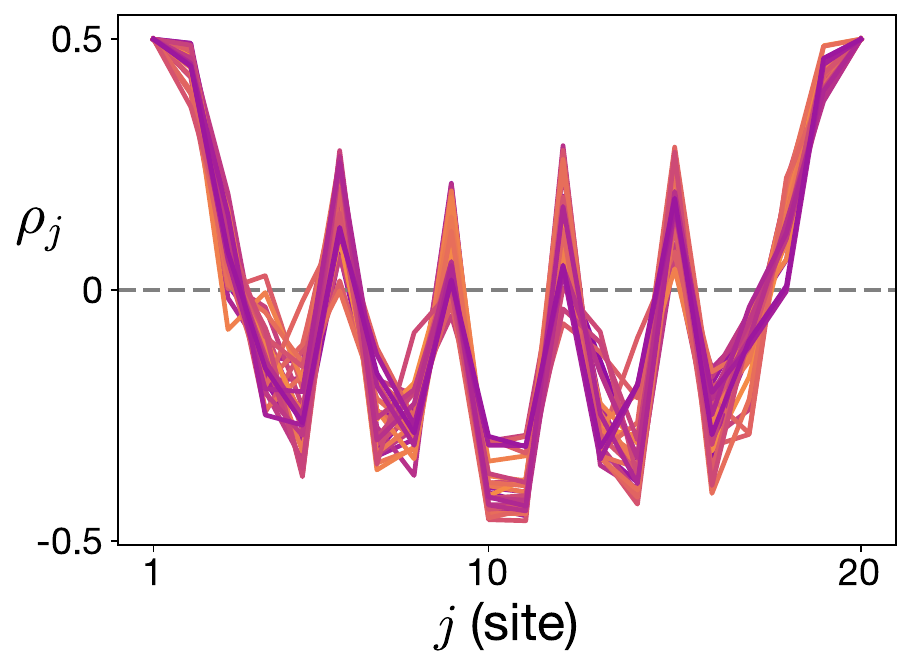}
\caption{Local charge distributions, $\rho_j$, for 40 mid-spectrum eigenstates of $H_\text{dip}(g_1, g_2)$ with $g_1=2, g_2=1$, for an open chain of length $L=20$. All eigenstates exhibit a charge distribution characteristic of a near-maximal quadrupole moment.} 
\end{figure}

\begin{figure}[h!]
\label{fig:S3}
\hspace*{1cm}
\includegraphics[width=0.9\linewidth]{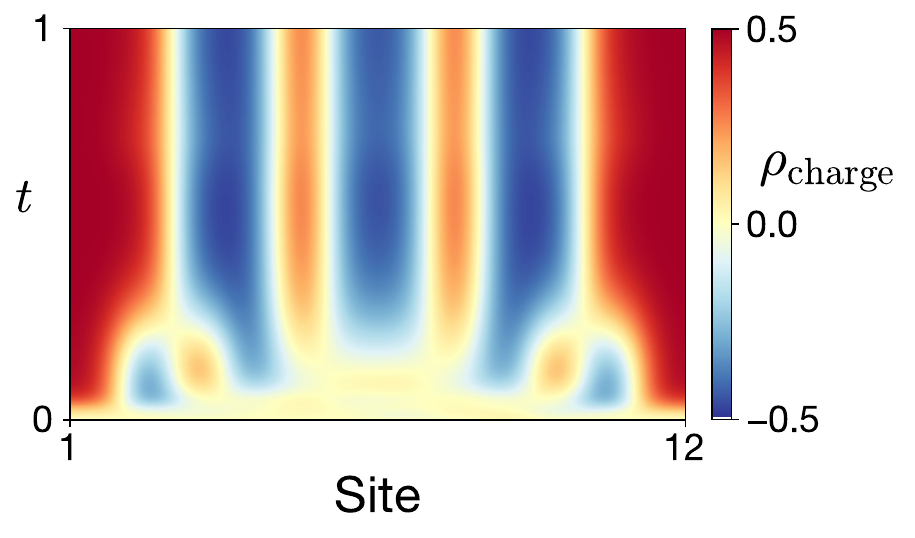}
\caption{Time evolution of a random state (averaged over 20 trials) using $H_\text{dip}(g_1, g_2)$ with $g_1=2, g_2=1$. The steady state exhibits a quadrupole moment of the charge.}
\end{figure}

For $g_1 = 2g_2$, on the other hand, the OBC spectrum is non longer real. Nevertheless, the key features of the dipole skin effect are the same as the model explored in the main text. In Fig.~\ref{fig:S1}, we show the spectrum of this system for an open chain of length $L=12$.  In Fig.~\ref{fig:S2}, we plot the local charge distribution for 40 mid-spectrum OBC eigenstates of this Hamiltonian for $L=20$, all of which have an extensive quadrupole moment. Lastly, the charge distribution after time evolving a random initial state with $H_\text{dip}(2g, g)$ is shown in Fig.~\ref{fig:S3}. As expected, the steady-state displays a symmetrical pattern of charge that reflects a maximal quadrupole moment.

\clearpage
\noindent
{\bf SM 2: Dipole skin effect in the presence of dipole-symmetry breaking terms}
\\
\indent
Starting from a dipole-conserving Hamiltonian such as Eq.~\eqref{eq:h4h5} and then adding charge hopping terms, as in the Hatano-Nelson model, explicitly breaks dipole conservation. As a result, it does not make sense to project the Hamiltonian into a sector of fixed dipole moment and look at the distribution of eigenstates. While one can still project into a sector of fixed charge (as it remains conserved), the resulting eigenstates will have very different charge distributions depending on their different dipole moments, obscuring the skin effect.

However, the dipole skin effect can still be seen in dynamics. In particular, we take a Hamiltonian with both dipole-hopping and charge-hopping terms:
\begin{equation}\label{eq:dipcharge}
H = H_\text{dip}(g_1, g_2) - \delta \sum_j (e^g c^\dag_j c_{j+1} + e^{-g} c^\dag_{j+1} c_j),
\end{equation}
where the second term is simply the Hatano-Nelson model, and $\delta$ controls the strength of dipole symmetry-breaking. When $\delta$ is smaller than the other scales in the system, the dipole skin effect continues to approximately hold, and a random initial state evolves under Eq.~\eqref{eq:dipcharge} to a state with a clear quadrupole moment. As $\delta$ becomes large and the charge-hopping dominates, the steady state instead exhibits a dipole moment, a signature of the charge skin effect dominating~over~the~dipole~skin~effect.
\begin{figure}[b!]
\label{fig:S4}
\includegraphics[width=0.9\linewidth]{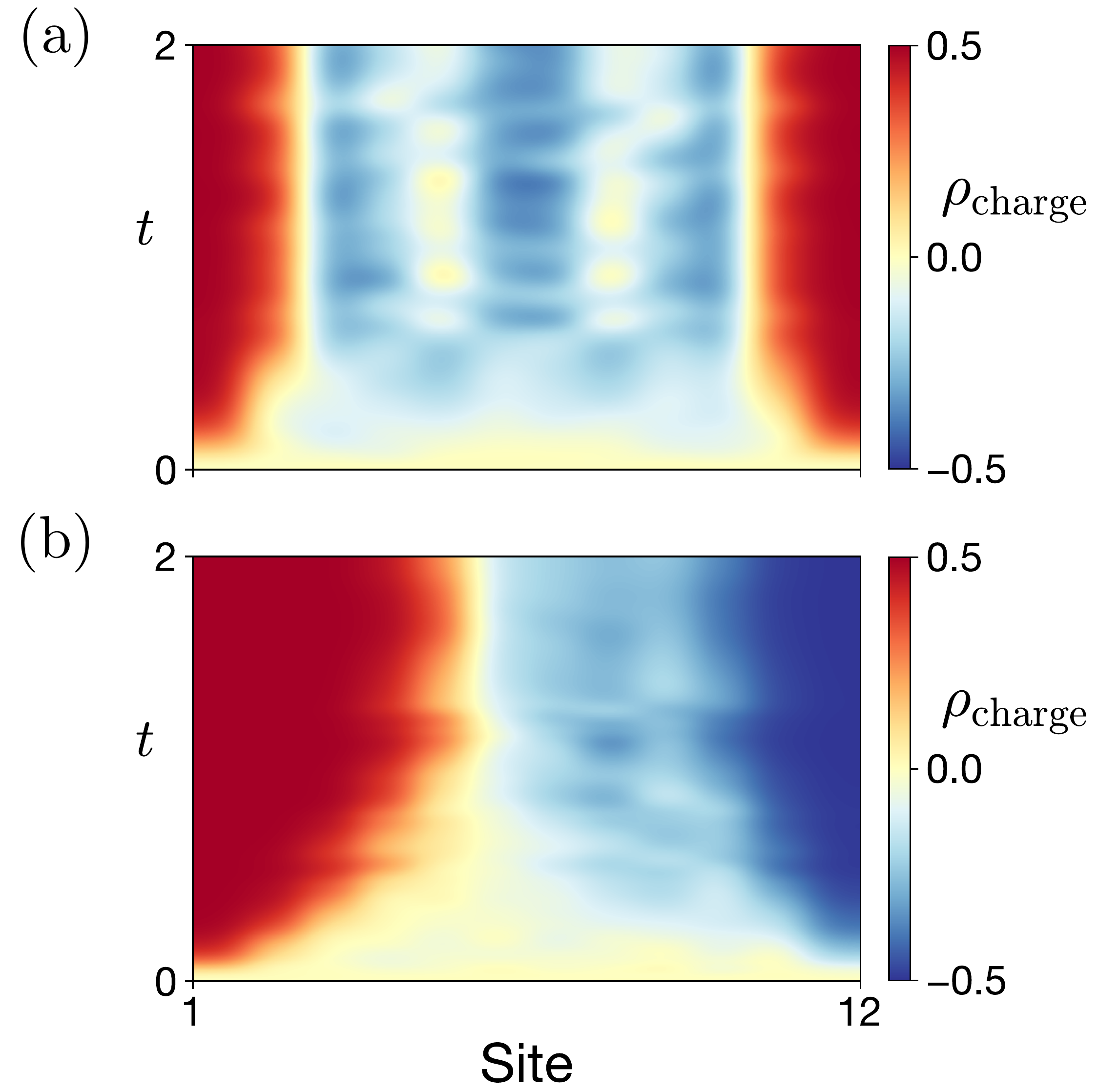}
\caption{Time evolution of a random state using a Hamiltonian with both charge and dipole hopping terms (Eq.~\eqref{eq:dipcharge} with $g_1=g_2=1$ and $g=1.5$), where $\delta$ controls the relative strength of charge-hopping over dipole hopping. (a) With small enough breaking of dipole symmetry ($\delta = 0.1$), the dipole skin effect approximately holds, (b) large dipole symmetry-breaking ($\delta = 10$) yields the charge skin effect.}
\end{figure}
These~two~scenarios~are~shown~in~Fig.~\ref{fig:S4}.

\end{document}